\documentstyle[seceq]{ptptex}

\newcommand{\be}{\begin{equation}}
\newcommand{\ee}{\end{equation}}
\newcommand{\bea}{\begin{eqnarray}}
\newcommand{\eea}{\end{eqnarray}}

\newtheorem{theorem}{Theorem}
\newtheorem{definition}{Definition}

\newtheorem{condition}{Condition}
\newtheorem{lemma}{Lemma}
\newtheorem{conjecture}{Conjecture}

\title{
Nature of singularities in gravitational collapse
}

\author{
Andrzej {\sc Kr\'olak}\footnote{E-mail address:
krolak@impan.gov.pl}
}

\inst{
Institute of Mathematics, Polish Academy of Sciences, \\
\'Sniadeckich 8, 00-950 Warsaw, Poland}

\abst{
We discuss several aspects of cosmic censorship hypothesis.
There is evidence both in favor of and against the hypothesis.
On one hand one can prove that cosmic censorship holds in several
special cases and on the other hand there are a number of special
solutions of Einstein equations in which it is violated. One way to
resolve cosmic censorship problem is to test it observationally. 
We point out several possibilities of such tests using present and
future instruments.
}

\begin{document}

\maketitle

\section{Introduction}

In a recent significant publication on the subject of space-time singularities
we read. 

"{\em One of the fundamental unanswered questions in the general 
theory of relativity is whether {\bf naked} singularities, that is
singular events which are visible 
from infinity, may form with positive probability in the process 
of gravitational
collapse. The conjecture that the answer to this question is in negative
has been called {\bf cosmic censorship}}" \cite{Ch1999}.
\footnote{It is interesting that physical concepts like
"gravitational collapse" 
and  physicists jargon like "naked singularities"
and "cosmic censorship" are getting into pure mathematics
literature.}

This clearly indicates that the fundamental question posed by 
Roger Penrose~\cite{P1969} that is cosmic censorship hypothesis remains
unresolved.
In this work I shall discuss some aspects of the cosmic censorship hypothesis
that are close to my own work on this problem. Moreover I shall 
point out possible observational tests of cosmic censorship. 
Other aspects of this conjecture are presented in a number of
contributions to these proceedings. Excellent exposition of
the cosmic censorship problem can be found in recent reviews
of Chru\'sciel \cite{Chr1992}, Clarke \cite{C1993}, Wald
\cite{W1997}, Hawking and Penrose \cite{HP1996}, Joshi \cite{J1997}, Singh
\cite{S1998}.

My review will be divided into three parts. First I shall present 
some theorems
using methods of global Lorentzian geometry restricting the occurrence of
naked singularities, next I shall discuss examples of naked singularities
in solutions of Einstein equations, and finally I shall point out a number 
of possibilities of observational verification of cosmic censorship. 

\section{Cosmic censorship: geometrical approach}

In 1969 \cite{P1969} Roger Penrose put forward a hypothesis that there exists
a "{\bf cosmic censor}" that forbids occurrence of "naked singularities".

In the language of Lorentzian geometry naked singularity is 
a {\em timelike ideal point} of the boundary of space-time.
The basic definitions and concepts of Lorentzian geometry can be found 
in the monograph of Hawking and Ellis \cite{HE1973}.

\begin{definition}
A future-directed (past-directed) causal curve $\lambda$ 
terminates in a timelike ideal point
if there exists a point p of space-time such that the chronological past
$I^-(\lambda) \subset I^-(p)$ (respectively chronological future $I^+(\lambda) \subset
I^+(p)$).
\end{definition}

Penrose showed \cite{P1979} that the absence of timelike ideal points is
equivalent to global hyperbolicity of space-time.

There are two versions of cosmic censorship: a strong one and a weak one.
The strong version says that space-time is globally hyperbolic.
The weak version says that the intersection of the causal future of a
partial Cauchy surface and the causal past of the boundary at infinity
is globally hyperbolic. Thus weak cosmic censorship says that there is 
no naked singularity outside the black hole event horizons.
For simplicity we shall not discuss possibility of violation 
of global hyperbolicity at infinity.
For example in the Oppenheimer-Snyder model \cite{OS1939} 
of gravitational collapse
the strong cosmic censorship holds whereas for Kerr 
space-time with $a^2 < m^2$
only the weak one is valid as the singularity that is hidden behind the event
horizon is timelike. 
Cosmic censorship does not say that there is no singularity visible
to observers. Clearly initial cosmological singularity is visible 
to all observers.
Rather it says that there exists an initial surface from which we can predict
evolution of the whole of space-time (strong version) or part of the space-time
outside black holes (weak version).

When space-time is not globally hyperbolic there is a partial Cauchy surface
$S$ such that Cauchy horizon $H^+(S)$ is not empty. Thus study of cosmic censorship
can be reduced to the study of the existence of Cauchy horizons.
A Cauchy horizon is a Lipschitz ($C^{1-}$) 3-dimensional manifold.
This, in turn, implies that Cauchy horizons are differentiable almost everywhere.  
Because they are differentiable except for a set of (three-dimensional) measure zero,
it seems that they have often been assumed to be smooth except for a
set that may be more or less neglected.  However, one must remember
in the above that: (1) the term "differentiable" refers to being
differentiable at a single point, and (2) sets of measure zero may be
quite widely distributed.
In fact Chru\'sciel and Galloway \cite{CG1998} have constructed 
examples of nowhere
differentiable Cauchy horizons and Budzy\'nski, Kondracki and this author
have shown~\cite{BBK1999} that a class of nowhere differentiable Cauchy
horizons is large.

We shall say that a Cauchy horizon $H$ is {\bf smooth} if
it {\em contains an open set G where its is $C^2$ and such that 
complement of $G$ in $H$ has measure zero}. Throughout the rest of 
this paper we shall assume that all Cauchy horizons are smooth.

The geometrical techniques to study the large-scale structure of space-time
developed by Geroch, Hawking, and Penrose use extensively an ordinary non-linear
equation of Riccati type known as Raychaudhuri-Newman-Penrose (RNP) equation.
Let $t$ be an affine parameter on a null geodesic $\lambda$, $K^a$ be components
of the tangent vector to $\lambda$, $\theta$ be expansion,
$\sigma$ be shear and $R_{ab}$ be components of the Ricci tensor.
then the RNP equation takes the form.
\be
\frac{d\theta}{dt} = -\frac{1}{2}\theta^2 -
2\sigma^2 - R_{ab}K^aK^b , \label{eq:RNP} \\
\ee
The quantity $\theta$ describes the expansion 
of congruences of null geodesics 
infinitesimally neighboring $\lambda$ and it is defined as  
$\theta = \frac{1}{A}\frac{dA}{dt}$ where $A$ is cross-section
of the congruence.

The great success of these techniques was the proof of the existence of
singularities (defined as incomplete causal geodesics) in general
space-times \cite{HP1970}. The techniques can be also used to study the
problem of cosmic censorship. We shall next review the results of this
and other authors in that direction.
We shall proceed as follows. 
We introduce a series of conditions on space-time and we discuss 
how each of them restricts the occurrence of naked singularities.

\begin{condition}[Null convergence condition]
We say that the null convergence condition holds if $R_{ab}K^aK^b \geq 0$ 
for all null vectors $K$.
\label{c:conv}
\end{condition}

By Einstein equations this condition is satisfied by all reasonable
classical matter models.

\begin{definition}
Let $S$ be a partial Cauchy surface.
A future Cauchy horizon $H^+(S)$ is compactly generated if
all its generators, when followed their past,
enter and remain in a compact subset $C$.
\label{d:cgen}
\end{definition}

The above class of Cauchy horizons has been introduced by Hawking 
\cite{H1992} to describe a situation in which a Cauchy horizon
arises as a result of causality violation rather than singularities or
timelike boundary at infinity.

\begin{theorem}[Hawking 1992\cite{H1992}]
If null convergence condition holds then a compactly generated Cauchy horizon 
that is non-compact cannot arise.
\end{theorem}

Thus under a very mild - from physical point of view - restriction 
on space-time a nontrivial class of Cauchy horizons is ruled out.

Let $R_{abcd}$ be components of Riemann tensor.
We say that an endless null geodesic $\gamma$ is
{\bf generic} if for some point $p$
on $\gamma$ \, $K^cK^dK_{[a}R_{b]cd[e}K_{f]} \neq 0$ where $K$ is a vector
tangent to $\gamma$ at $p$.

\begin{condition}[Generic condition]
All endless null geodesics in space-time are generic.
\label{c:gen}
\end{condition}

This condition means that every null geodesic encounters 
some curvature that is not specially aligned with geodesic.

\begin{theorem}
If null convergence condition holds and one of the null geodesic 
generators of a Cauchy horizon $H$ is generic then $H$ cannot 
be compact.
\end{theorem}

The above result is a consequence of the properties of the compact future 
Cauchy horizon $H$ given by the following lemmas.

\begin{lemma}[Hawking and Ellis 1973\cite{HE1973}]
The null geodesic generators of $H$ are past complete.
\label{l:hcom}
\end{lemma}

\begin{lemma}[Hawking and Ellis 1973\cite{HE1973}]
Let null convergence condition hold.
Then the expansion $\theta$ and the shear $\sigma$ of null geodesic
generators
of $H$ are zero.
\label{l:hexp}
\end{lemma}

\begin{lemma}[Borde 1984\cite{B1984}, Hawking 1992\cite{H1992}]
There exists an endless null geodesic generator of $H$.
\label{l:hend}
\end{lemma}

\begin{lemma}[Beem and Kr\'olak 1998\cite{BK1998}]
Let null convergence condition hold. Then
all null geodesic generators of $H$ are endless.
\label{l:haend}
\end{lemma}

From Lemmas \ref{l:hcom}, \ref{l:hexp}, \ref{l:hend} 
it already follows that
existence of compact Cauchy horizons is incompatible with
generic conditions. This is a result of Borde \cite{B1984}.
By Lemma~\ref{l:haend} a compact Cauchy horizon cannot contain
even one generic generator. 
Thus we see that under Conditions \ref{c:conv} and \ref{c:gen} 
modulo certain differentiability assumptions compact Cauchy horizons 
are ruled out.
It is interesting to note that Conditions \ref{c:conv} and \ref{c:gen} 
are one of the assumptions of the Hawking-Penrose singularity
theorem~\cite{HP1970}.

\bigskip
{\em Remark}

Lemmas \ref{l:hcom} and \ref{l:hend} apply to the case of a compactly
generated
Cauchy horizon that is not necessarily compact. The past-complete and
future-endless generators of $H$ are then contained in the compact set $C$
in Definition~\ref{d:cgen}~\cite{H1992}.
\bigskip

Let $\lambda$ be a past endless achronal null geodesic. We say that $\lambda$ 
is {\bf past focusing} if there exists
a point $q$ on $\lambda$ such that the expansion $\theta$ of the congruence
of past-directed null geodesics originating from $q$ and infinitesimally
neighboring to $\lambda$ becomes negative at some point $p$ on $\lambda$.
By time inverse of the above we define {\bf future focusing} null
geodesics.

\begin{condition}[Strong null convergence condition]
We say that the strong null convergence condition holds if every past
(future) endless achronal null geodesic terminating 
in the timelike ideal point of the boundary of space-time 
is past focusing (respectively future focusing). 
\end{condition}

\begin{theorem}[Kr\'olak 1987\cite{K1987}, 
Kr\'olak and Rudnicki 1993\cite{KR1992}]
Let $S$ be a partial Cauchy surface with an asymptotically simple past 
and let the Cauchy horizon $H^+(S)$ be non-empty.
If null convergence condition holds and a generator $\lambda$ of
$H^+(S)$ is past-focusing then the set
$C := \overline{I^-(q) \cap S}$ must be compact for some point  $q \in
\lambda$.
\label{t:hrn}
\end{theorem}

This result can be interpreted as a topological instability of
Reissner-Nordstr\"om and Kerr type Cauchy horizons for which the
intersection $C := \overline{I^-(q) \cap S}$ for every point q on the
Cauchy horizon is non-compact. By results of Tipler \cite{Ta1977}
it follows that for Reissner-Nordstr\"om space-time 
an arbitrary small amount of outgoing spherically
symmetric radiation in some compact neighborhood of the intersection
of the event horizon with a spacelike hypersurface will cause
the past strong null convergence condition to be satisfied on
the Cauchy horizon. Studies of perturbations of Reissner-Nordstr\"om
space-time have established not only instability
of these horizons but also in a remarkably great detail the
structure of the singularity of the perturbed space-time
(see contribution of Brady in this volume and also recent
paper by Burko~\cite{B1999}.
The advantage of the above result is that it applies not only 
to a Cauchy horizon of a particular highly symmetric solution
of Einstein equations but also to all Cauchy horizons of a certain
topological type without any symmetries of space-time.

\begin{definition}
Space-time is maximal null pseudoconvex if and only if for each compact
set $K$ there
exists a compact set $K'$ such that each maximal null geodesic segment 
with both endpoints in $K$ must have its image in $K'$.
\end{definition}

Null pseudoconvexity (a condition marginally stronger than the above) 
together with condition of null geodesic disprisonment 
which follows from strong causality 
condition imply global solvability of inhomogeneous wave equations
\cite{BP1987}.  Thus these conditions play a similar role in the theory
of partial differential equations as global hyperbolicity but are
weaker. It was demonstrated that pseudoconvexity could be used in place
of global hyperbolicity in study of Lorentzian geometry.
For example it implies equality of lower and upper Hausdorff limits for
sequences of geodesics~\cite{BP1987}.
Intuitively, one may think of pseudoconvex space-times as those 
failing to have any "interior" points missing.

\begin{theorem}[Beem and Kr\'olak 1992\cite{BK1992}]
Let space-time be strongly causal. If both the null convergence
and the strong null convergence conditions hold then space-time
is maximally null pseudoconvex.
\end{theorem}

The significance of the above theorem for cosmic censorship is that
it proves a causality condition that can in some cases be used
instead of global hyperbolicity to ensure predictability of space-time.

\begin{condition}[Trapped surface condition]
The trapped surface condition holds if
for every future-incomplete non-spacelike geodesic $\lambda$,
$\lambda \in {\it int}D(S)$ where $S$ is a regular
partial Cauchy surface there exists a trapped surface ${\cal T}$
such that $J^+({\cal T}) \cap \lambda \neq \emptyset$.
\end{condition}

Hawking-Penrose singularity theorem says that when space-time is strongly 
causal and  conditions \ref{c:conv} and \ref{c:gen} hold then the 
existence of a trapped surface implies existence of an incomplete 
causal geodesic. Thus the trapped surface
condition essentially says that the inverse of singularity theorem holds.

\begin{definition}
Let $(M,g)$ be a weakly asymptotically simple and empty space-time.
A partial Cauchy surface $S$ in $M$ is said to be regular
if the following conditions hold.

1. $\overline{D^+(S,\overline{M})} \cap \lambda \neq \emptyset$   
   for all generators $\lambda$ of ${\cal J}^+$.

2. $S$ has an asymptotically simple past.

3. If $H^+(S) \neq \emptyset$  then for every past-incomplete null geodesic
generator $\gamma$ of $H^+(S)$ there exists a point $p \in \gamma \cap
H^+(S)$
such that a set  $\overline{I^-(p) \cap S}$ is compact.
\label{Def:reg}
\end{definition}

The purpose of the above definition is to describe in a geometrical way
what the regular initial data are i.e. to ensure that the break down of
prediction does not
arise from a bad choice of the initial surface. The definition originated from 
the concept of partial asymptotic predictability introduced by Tipler \cite{T1976}
that requires that at least some structure of ${\cal J}^+$ can be predicted
from initial data. Condition 3 eliminates Cauchy horizons of the
Reissner-Nordstr\"om type. If we assume that every generator of
a Cauchy horizon is past-focusing then this condition follows from
Theorem \ref{t:hrn}.

\begin{theorem}[Kr\'olak 1986\cite{K1986}, Kr\'olak and
Rudnicki 1993\cite{KR1993}]
Let space-time be weakly asymptotically simple and empty.
If space-time contains a regular partial Cauchy surface $S$
and if null convergence, generic, 
and trapped surface conditions hold then
space-time is future asymptoticaly predictable from $S$.
\end{theorem}

The above theorem proves predictability of space-time under the
trapped surface condition which is very strong. 
Nevertheless the theorem is non trivial and regularity
required by Definition \ref{Def:reg} is essential for its validity.

The final conclusion of this section is that theorems based on methods of
Lorentzian geometry do not restrict the occurrence of Cauchy horizons 
to such a degree that we can accept the cosmic censorship principle.

\section{Naked singularities in gravitational collapse}

In this section we shall discuss various instances of occurrence of naked
singularities in solutions of Einstein equations.

\subsection{Shell-crossing singularities}

The shell-crossing singularities are the earliest examples of naked
singularities in gravitational collapse. They were first found and studied
in detail in Lemaitre-Tolman-Bondi (LTB) space-times
representing spherically symmetric collapse of dust~\cite{YSM1973}.
Even though they arise from regular initial data they were never thought
to constitute a serious counterexample to cosmic censorship.
One reason is that metric admits a $C^0$ extension through such
singularities \cite{Ch1984}. Other reasons emerge during the following
discussion.

\subsection{Strong curvature singularities}

Before we introduce another type of naked singularities let us
recall the concept of strong curvature singularity. This is an idea
of gravitationally strong singularity that destroys by crushing or stretching
any object that falls into it. The idea was first defined in precise
geometrical terms by Tipler \cite{T1977} and then two kinds of
strong curvature singularities emerged a {\em strong curvature singularity}
that crushes all volume elements defined by Jacobi fields to zero
and a {\em limiting focusing singularity} that causes all volume elements
to decrease. It turned out those strong curvature singularities
can be characterized by non-integrability of certain components
of curvature tensor along causal geodesics fall into
them~\cite{Ta1977,CK1985}. 
A singularity that satisfies either strong curvature condition
or limiting focusing condition is past or future focusing.
A classification of strong curvature singularities
in the case of spherical symmetry was given by Nolan \cite{N1998}.

The following conjecture emerged~\cite{T1977,K1978}.

\begin{conjecture}
If all singularities that arise in space-time are of strong curvature
then cosmic censorship holds.
\label{c:sch}
\end{conjecture}
 
\subsection{Shell-focusing singularities}

Shell focusing singularities were discovered 
by Eardley and Smarr \cite{ES1979}
in spherical collapse of dust matter. 
They were studied in detail by Christodoulou \cite{Ch1984} and Newman
\cite{RPAC1986}. 
These singularities have the property
that they arise from the evolution of central degenerate shell of matter
and they have zero mass.
It was Newman \cite{RPAC1986} who has shown that shell-focusing
singularities
were limiting focusing singularities and consequently the Conjecture 
\ref{c:sch} put
forward by Tipler and this author turned out to be false
and attempts to proof cosmic censorship on that basis failed.
Nevertheless by results presented in previous section, 
some of which originated from the attempts to prove
Conjecture \ref{c:sch}, the occurrence of naked singularities 
in general space-times is somewhat constrained.  

Newman also showed that shell-crossing singularities do
not satisfy either strong curvature or limiting focusing
condition. Consequently shell-crossing singularities are
gravitationally weak and integrable.

Subsequently Joshi and his school~\cite{J1993,JD1992,SJ1996,JJS1996}
found many examples 
of shell-focusing singularities that were either strong curvature 
or limiting focusing. 
Ori and Piran \cite{OP1987} discovered shell-focusing singularities
in spherically symmetric self-similar collapse of perfect fluid and the
singularities were shown by Lake \cite{L1988} to be of strong curvature 
type.
Lake \cite{L1991} gave the first examples of non self-similar
shell-focusing singularities and Harada
\cite{H1998} showed that shell-focusing singularities occur for perfect
fluid with a sufficiently soft equation of state. 
Joshi and this author \cite{JK1996}
showed that strong curvature shell-focusing singularities occur in 
Szekeres space-times that do not have any Killing vectors. 

One natural question that arises is whether generic perturbations
will destroy naked singularities. Iguchi, Nakao, and Harada \cite{INH1998}
found that within linear perturbations odd-parity gravitational waves
do not destroy the Cauchy horizon forming as a result of naked
shell-focusing singularity in spherically symmetric dust collapse.

\subsection{Scalar field naked singularities}

Naked singularities were found numerically in critical collapse
space-times \cite{Cz1993}. However, since the naked singularity 
is realized for a specific solution in the one parameter family, it is a
subset of measure zero. 
Exact solutions for the case of massless scalar field
containing naked singularities were discovered by Roberts
\cite{R1989,R1996}.
Christodoulou made a complete study of gravitational collapse 
of scalar field. Among other things he proved~\cite{Ch1994} that
naked singularities
occur in gravitational collapse of self-similar scalar field.
However he showed that the {\em set ${\cal E}$ in the space of initial
data leading
to formation of naked singularities has positive codimension}.
Thus at least for this model example we can say that cosmic censorship
holds.

\subsection{Collapse of collisionless dust}

One line of thought is that in order that cosmic censorship holds
the matter model must be physically realistic.
Rendall \cite{R1992} proposed that an appropriate set of
physically realistic set of equations is Einstein-Vlasov system.
He pointed out a very appealing analogy with Poisson-Vlasov
set of equations which was proven to be
generically free of singularities whereas Newton's equations
describing evolution of dust do exhibit singularities.
Rendall pointed out that velocity dispersion
present in Einstein-Vlasov case can dissolve naked singularity 
\cite{R1992} in the same way as in the  Poisson-Vlasov case.
Shapiro and Teukolsky \cite{ST1991} found numerically singularities 
without formation of apparent horizon in axially symmetric collapse 
of collisionless cloud of particles.
The set of equations evolved by Shapiro and Teukolsky is a special
case of Einstein-Vlasov system but they assumed
that particles were initially at rest. This means that in their solution
there was no velocity dispersion. 
Recently it was shown by Harada, Iguchi and Nakao \cite{HIN1998}
that for a spherical cloud of counterrotating  particles the formation
of shell-focusing singularity is prevented. As counterrotation can be
interpreted as a simple model of velocity dispersion this result
supports the line of attack on cosmic censorship proposed by Rendall.

Rein and Rendall proved~\cite{RR1992} global solvability of
Einstein-Vlasov system for small
initial data. One significance of this result is that in Eistein-Vlasov
system there cannot be shell-crossing singularities present in
LTB space-time because in that case they can occur for arbitrarily
small initial data.

\section{Observational verification of cosmic censorship}

An insight into the nature of the final state of gravitational collapse can be 
provided by measurement of the dimensionless quantity
\be
\chi := \frac{a}{m}
\ee
where $m$ is mass and $a$ is spin angular momentum per unit mass. 
For Kerr space-time with $\chi < 1$ weak cosmic censorship holds
whereas for $\chi > 1$ it is violated.
I shall discuss a number of possibilities to measure the parameter
$\chi$.

\subsection{Pulsar observations}

Observations of radio pulsars proved to be a powerful
tool for studying compact objects. Pulsar in orbit with an object more
compact than a neutron star is yet to be discovered. Once
it is found the ratio $a/m$ can be measured form the timing model.
Two effects were considered.
One is the time delay $\Delta_{FD}$ in the propagation of electromagnetic
impulses from the pulsar due to frame dragging \cite{NPS1991,LW1997}.
The maximum of this delay is given by \cite{WK1998}
\be
\mbox{max}(\Delta_{FD}) \approx \frac{0.8\times10^{-3}}{|\cos i|}
\left(\frac{P_b}{\mbox{1 day}}\right)^{-2/3}\left(\frac{m_{\odot}}{10}\right)^{5/3}
\chi\cos\lambda \, [\mu s], 
\ee
where $\pi - i$ is inclination of the orbit to the line-of-sight,
$P_b$ is orbital period, $m_{\odot}$ is pulsar companion mass in units of solar mass,
and $\lambda$ is the angle between the companion spin vector and the line-of-sight. 
It was pointed out by Wex and Kopeikin \cite{WK1998} that the measurement 
of the frame-dragging effect is complicated by a competing effect of 
the bending delay.
Instead they found that the spin-induced precession will have influence 
on observable quantities of the timing
model. In particular measurement of projected semi-major axis
of the pulsar orbit and periastron advance will be perturbed by
a factor proportional to $\chi$. They have shown that one can determine
parameter $\chi$ accurately within a reasonable span of observations.

\subsection{Gravitational-wave observations} 

The gravitational wave signal from a binary system will carry 
information about the spin of the members of the system.
Spin-orbit and spin-spin interactions enter respectively 1.5 PN 
and 2 PN corrections to the phase of the gravitational
signal~\cite{KWW1993}. From observations of gravitational-waves by laser
interferometric detectors one can in principle determine masses 
and parameter $\chi$ of the members of the binary
system~\cite{PW1995,KKS1995}. Simplified analysis shows that one can estimate
parameter $\chi$ most accurately for a massive companion 
of a typical neutron star.
For advanced LIGO interferometer one gets relative rms error
$\frac{\Delta\chi}{\chi} \sim \frac{10}{\chi}\%$ at 200Mpc
for a fiducial binary ($m_{NS} = 1.4M_{\odot}, m = 10M_{\odot}$)
where $m_{NS}$ is the mass of a neutron star. The absolute rms
error in the estimation of $\chi$ is almost independent of its
value and the relative error is the smaller the larger the ratio
$a/m$.

\subsection{Measurements of X-ray binaries spectra}

The existence and properties of very compact objects like black holes
can be investigated through their interaction with other bodies.
For example an accretion disc can form around a compact object.
The X-ray radiation emitted during accretion can be detected 
by satellites.
Both binaries with a central object of the mass of a few tenths of solar
masses and active galactic nuclei (AGN) powered by compact objects of
masses equal to masses of galaxies can be sources of X-ray radiation.

In the case of stellar sized compact objects in the spectra of low
mass X-ray binaries (LMXB) one observes quasiperiodic oscillations (QPO).
These oscillations can be explained by relativistic effects in Kerr
solutions.
One possibility is that they are due to oscillations of accretion
disc. The normal modes eigenfrequencies of accretion disk 
depend on mass and $\chi$ parameter of the spinning central object
\cite{NL1998} and can in principle be estimated form QPO frequencies.
The other possibility is that QPOs can be directly related to precession
frequency of the accretion disc through Lens-Thirring effect
\cite{CCZ1998}. 
\be
\Omega_p = \frac{2\chi}{r^3},
\ee
where $\Omega_p$ is precession frequency of the accretion disc and
$r$ is radius. Thus parameter $\chi$ can be determined.

In the case of supermassive objects the best tests of cosmic censorship
may come from measurements of
iron $K_{\alpha}$ line profiles from Seyfert type 1 galaxies.
The X-ray spectra of these AGNs show evidence for line emission peaking at
a rest energy of 6.4 keV. This is thought to be due to a fluorescence line 
from the K-shell of iron.
The lines are extremely broad and have a strong asymmetry to the red. 
This emission is thought to originate from the innermost regions of an
accretion disk around a supermassive central compact object.
These lines provide the means to probe the immediate environment 
of a black hole. The most prominent example of such a line comes
from the galaxy MCG-6-30-15~\cite{Tal1995,Ial1997,Gal1998}.
The model of the line depends on $a/m$ of the central object
and several other parameters: inner and outer radii of the accretion disc,
inclination angle and emissivity index (Fabian et al.~\cite{FRSW1989} and
see Fanton et al.~\cite{FCFC1997} for a systematic derivation).
In particular the full width at zero intensity (FWZI) $\Delta$ depends
on $\chi$ and we have $\Delta > 4/\sqrt{3}$ for $\chi > 1$.
From current data it is not possible to conclude whether $\chi$ is $0$
or $1$ \cite{Nal1999}. The recently launched CHANDRA X-ray satellite 
should provide more accurate measurements.

\subsection{Evolution of $a/m$ due to accretion}

When central object accretes matter from the surrounding disc
both its mass and angular momentum change. The resulting evolution 
of ratio $a/m$ has been studied for the case of $a/m < 1$ by
Bardeen \cite{B1970} and Thorne \cite{T1974} and for the
case of $a/m > 1$ by de Felice \cite{deF1978}.
Bardeen found that an initially non-rotating black hole would get spun up
to extreme Kerr black hole ($a/m = 1$). Thorne \cite{T1974} showed that
because 
capture cross section is greater for negative angular momentum photons 
than for positive angular momentum photons
black hole will be spun up to $a/m \simeq 0.998$. The case when initial 
$a/m$ of the central compact object is greater than $1$ was considered
by de Felice~\cite{deF1978} who showed that in this case $a/m$ decreases.
Because of
the effect studied by Thorne will also operate in this case the
compact object should get spun down to $a/m \simeq 0.998$.
Even though this calculation based on Kerr solution indicates
(classical) instability of a naked singularity there are
other physical effects like magnetic field that may change
the above picture. De Felice~\cite{deF1978} calculated that for the
range $1 < \chi < 4/3 \sqrt{2/3}$ the innermost stable orbits
of Kerr solution have negative energy and consequently 
the efficiency of the photon emission can be more than 100\%.
Hence an extremely compact object spinning down from the initial $a/m$
greater than $1$ may be a very powerful
source of energy. An early numerical work of Nakamura and Sato
\cite{NS1981} shows that for some equations of state as a result of
collapse of a rotating star with initial value of parameter $\chi > 1$
a jet forms. It was suggested by Chakravarti and Joshi \cite{ChJ1992}
that naked singularities may be sources of gamma rays bursts.

It is interesting to note that all the tests of cosmic censorship
discussed above do not require any new expensive instrument but
they can use existing and planned observational projects. 
One is only required in the analysis of the data
to take a sufficiently large parameter space and not assume
beforehand that $a/m$ of a compact object is necessarily less than $1$.

\section{Conclusion}

The results presented in our review show that there is evidence
both against and for the existence of cosmic censor. 
It appears that analytic, numerical,
and observational techniques that we have are not yet sufficiently
refined to tackle this problem. Thus cosmic censorship often
referred to as the most fundamental unsolved problem of general 
relativity remains a challenge for the next - 21st century.

\section{*Acknowledgments}

The Polish Committee for Scientific Research supported this work 
through grant 2~P03B~130~16.

\end{document}